\documentstyle[graphicx,float]{article}

\begin{document}
\title{Comparing entanglement and identity effects in tunnelling}
\date{}
\author{Pedro Sancho \\ Centro de L\'aseres Pulsados CLPU \\ Parque Cient\'{\i}fico, 37085 Villamayor, Salamanca, Spain}
\maketitle
\begin{abstract}
The probability of simultaneous tunnelling of two particles is modified when the system is in a non-separable state, either entangled or symmetrised. We compare both effects in the rectangular potential barrier by evaluating the transmission rates in superposition states. For large momenta, their simultaneous presence greatly changes the form of the transmission rates. The joint effects are much larger than the superposition ones. Moreover, there are significant differences between bosons and fermions. We present an unified view of the combined effects as a quantum interference phenomenon. The analysis also illustrates a novel aspect of exclusion in entangled systems, the existence of superposition states one of whose terms is forbidden by Pauli's principle.
\end{abstract}

\section{Introduction.}

The relation between the two types of non-separability in quantum mechanics, entanglement and identity, is a subject of increasing interest. On the one hand, the problem of the definition of entanglement measures for identical particles has not been completely settled yet \cite{tri}. On the other hand, several schemes have been proposed to convert the non-separability of systems of identical particles into useful entanglement \cite{bra,ple,lof}.

A related topic is the comparison of both effects and the interplay between them when they act at once. In \cite{yap,yej} light absorption by identical atoms in multi-particle superposition states was studied, showing a rich variety of novel behaviours, such as the inhibition of the dependence on overlapping for fermions for some states or the  cancellation  of the entanglement effects for some values of the parameters. In general, sharp differences between bosons and fermions were observed. 

In this paper we want to go deeper into the problem by studying another arrangement that can be treated in a first approximation in an analytical way, the tunnelling of a barrier by pairs of identical particles. We evaluate the probability of simultaneous tunnelling of the two particles in the archetypal rectangular potential barrier \cite{gri,gal}. The simultaneous transmission rates show large deviations with respect to those of product states and distinguishable particles. 

The role of  entanglement in tunnelling processes has been previously considered in the literature, specially in the context of Bose-Einstein condensation and of multi-boson systems in optical lattices. For instance, in \cite{ng} the correlated tunnelling in condensates of two interacting species leads to entanglement between two macroscopic systems. Similarly, in \cite{alc} the entanglement dynamics in macroscopic tunnelling of condensates has been analyzed in detail. However, our focus here is completely different, we are only interested in two-particle systems. Some aspects of our approach are also connected with other studies in the literature. For instance, in the Hong-Ou-Mandel arrangement \cite{hom}, the rate of simultaneous detection in the exit paths depends on the degree of entanglement between the photons impinging on the beam splitter \cite{lou}. Also, in a related context, interferometry in multi-particle superposition states have been extensively studied \cite{zei}.  

The comparison of the detection patterns of superposition and mixed states in multi-particle two-slit interference provides a method to analyze multi-particle coherence in the problem \cite{yed}. The comparison can also be used to study the role of multi-particle coherence in tunnelling processes. As for pure states of distinguishable particles, entanglement is equivalent to coherent multi-particle superposition, this is a good characterization of entanglement effects. On the other hand, for identical particles, it provides a measure of the importance of multi-particle superposition in the presence of exchange effects.

Another manifestation of the interplay between multi-particle superposition and exchange effects is the existence of excluded states not related to Pauli's principle \cite{yap,yej}. We shall find here another aspect of exclusion associated with the presence of entanglement. There are states forbidden by Pauli's principle \cite{pau} that, however, can be present in a superposition. This is a multi-particle coherence effect. It highlights in a drastic way how entanglement can modify the behaviour of multi-fermion systems. 

\section{The arrangement.}

Two particles, distinguishable or identical, impinge simultaneously on a rectangular potential barrier of width $d$ and height $V_0$. The energy of the two particles is less than $V_o$. Because of the tunnel effect the particles have a non-zero probability of penetrating the barrier and to be detected on the classically forbidden region behind the obstacle. We want evaluate the probability of simultaneous tunnelling of the two particles.

In order to analyze this probability in the different cases signaled in the Introduction we must consider different pure and mixed initial states of distinguishable and identical particles. In the distinguishable case we deal with the superposition state
\begin{equation}
|\Psi>=N(a|\psi>_A|\phi>_B+b|\varphi>_A|\chi>_B)
\end{equation}
and the mixed one represented by the density matrix
\begin{equation}
\rho _{\Psi} =|a|^2 |\Psi_a><\Psi_a| + |b|^2 |\Psi_b><\Psi_b| 
\end{equation}
with 
\begin{equation}
|\Psi_a>=|\psi>_A|\phi>_B \; , \; |\Psi_b>=|\varphi>_A|\chi>_B
\end{equation}
the two product states entering the superposition. In these expressions the subscripts $A$ and $B$ denote the two particles, and $a$ and $b$ are the coefficients of the superposition, obeying the relation $|a|^2+|b|^2=1$. The four one-particle states are normalized, and the normalization coefficient of the superposition is
\begin{equation}
N=(1+2Re(a^*b <\psi|\varphi><\phi|\chi>))^{-1/2}
\end{equation}
Note that the one-particle states of $A$ and $B$ must be non-orthogonal in order to impinge on the same potential barrier.

For identical particles we have 
\begin{equation}
|\Phi>={\cal N}\{ a(|\psi>_1|\phi>_2 \pm |\phi>_1|\psi>_2)   +b(|\varphi>_1|\chi>_2 \pm |\chi>_1|\varphi>_2)\}
\end{equation}
for the superposition, and
\begin{equation}
\rho _{\Phi} =|a|^2 |\Phi_a><\Phi_a| + |b|^2 |\Phi_b><\Phi_b| 
\end{equation}
for the mixture, with 
\begin{equation} 
|\Phi_a>={\cal N}_a(|\psi>_1|\phi>_2 \pm |\phi>_1|\psi>_2) 
\end{equation}
and
\begin{equation}
|\Phi_b>={\cal N}_b(|\varphi>_1|\chi>_2 \pm |\chi>_1|\varphi>_2)
\end{equation}
The subscripts have been substituted by $1$ and $2$ to emphasize that now we are dealing with non-distinguishable particles. In the double sign expressions $\pm$, the upper one holds for bosons and the lower one for fermions. The normalization factors are
\begin{equation}
{\cal N}_a =(2 \pm 2|<\psi |\phi>|^2)^{-1/2} \; , \; {\cal N}_b =(2 \pm 2|<\varphi |\chi>|^2)^{-1/2}
\end{equation}
and 
\begin{eqnarray}
{\cal N}^{-2}=|a|^2 {\cal N}_a^{-2}+ |b|^2 {\cal N}_b^{-2}+ \nonumber \\
4Re(a^*b\{ <\psi|\varphi><\phi|\chi> \pm <\psi|\chi><\phi|\varphi> \})
\end{eqnarray}
In order to have exchange effects some of the one-particle states of the two particles must overlap, that is, $<\psi|\phi> \neq 0$, $<\psi|\chi> \neq 0$, $<\varphi|\chi> \neq 0$ or $<\varphi|\phi> \neq 0$.
 
The initial one-particle wave functions are multi-mode packets sharply picked around a central value \cite{pr}:
\begin{equation}
\psi (x,t)=(2\pi \hbar)^{-1/2} \int dp_* f(p_*) \exp (i(p_*x - Et )/\hbar)
\end{equation}
with the Gaussian mode distribution given by
\begin{equation}
f(p_*)=\frac{2^{1/4}}{\pi^{1/4}P^{1/2}} \exp (-(p_*-p)^2/P^2)
\end{equation}
with $P$ determining the width of the distribution, $p_*$ the momentum variable, $p$ the central value, $E$ the energy, and $x$ the spatial coordinate of the one-dimensional arrangement. $f(p_*)$ is the wave function in the momentum representation and, consequently, it is normalized to unity.

\section{Evolution and probabilities.}

After interacting with the potential barrier the evolution of the one-particle states is
\begin{equation}
|\psi> \rightarrow R |\psi_R>+T|\psi _T>
\label{eq:tre}
\end{equation}  
where $R$ and $T$ are the reflection and transmission coefficients. Similarly, the subscripts denote the reflected and transmitted states. 

Using Eq. (\ref{eq:tre}) and the similar expressions for the other one-particle states, the final state of the superposition of distinguishable particles is 
\begin{equation}
|\Psi _f>= N(aT_A(p)T_B(q)|\psi_T>_A |\phi _T>_B + bT_A(\bar{p})T_B(\bar{q})|\varphi_T>_A |\chi _T>_B   )+\cdots
\end{equation} 
$T_A(p)$ denotes the transmission coefficient for particle $A$ with central momentum $p$. The rest of terms in $\Psi _f$ contain reflected states. As we are only interested into double transmission its explicit form is irrelevant here and is not included in the above equation.

All the contributions to double absorption are contained in the explicitly written term of the r. h. s. of the above equation. Consequently, the probability amplitude of double absorption is given by the matrix element of the state corresponding to that normalized term with the final state, that is, $<2|\Psi _f>$ with 
\begin{equation}
|2> =N_T(aT_A(p)T_B(q)|\psi_T>_A |\phi _T>_B + bT_A(\bar{p})T_B(\bar{q})|\varphi_T>_A |\chi _T>_B   )
\end{equation}
and
\begin{eqnarray}
N_T^{-2}=|aT_A(p)T_B(q)|^2+|bT_A(\bar{p})T_B(\bar{q})|^2+ \nonumber \\
2Re(a^*b T_A^*(p)T_B^*(q)T_A(\bar{p})T_B(\bar{q})<\psi_T|\varphi_T><\phi_T|\chi_T>)
\end{eqnarray}
The addition of the two contributions in $|2>$ can also be understood from the point of view of the physical alternatives for double transmission. The two alternatives are $|\psi>|\phi> \rightarrow |\psi_T>|\phi_T>$ and  $|\varphi>|\chi> \rightarrow |\varphi_T>|\chi_T>$. As these alternatives are indistinguishable their probability amplitudes add (with the adequate weights, as done in \cite{yap}). This is equivalent to evaluate the total probability amplitude with the state $|2>$.

Then, the probability of double transmission for the superposition of distinguishable particles  is 
\begin{equation}
P_{dis}^{sup}=|<2|\Psi _f>|^2=|<2|N_T^{-1}N|2>|^2=N^2/N_T^2
\end{equation}
Similarly, for the initial product states $\Psi_a$ and $\Psi_b$ we have
\begin{equation}
P_{dis}^a=|T_A(p)T_B(q)|^2 \; , \;  P_{dis}^b=|T_A(\bar{p})T_B(\bar{q})|^2
\end{equation}
and for the mixture
\begin{equation}
P_{dis}^{mix}=|a|^2|T_A(p)T_B(q)|^2 +|b|^2|T_A(\bar{p})T_B(\bar{q})|^2
\end{equation}
In the case of identical particles we consider first the non superposed states. For $\Phi _a$ the interaction with the barrier leads to an expression of the type $T(p)T(q){\cal N}_a(|\psi_T>_1|\phi _T>_2 \pm |\phi_T>_1|\psi_T>_2 )+\cdots $. All the contributions to double transmission are now contained in the state ${\cal N}_T^a(|\psi_T>_1|\phi _T>_2 \pm |\phi_T>_1|\psi_T>_2 ) $, with ${\cal N}_T^a=(2\pm 2|<\psi_T|\phi_T>|^2)^{-1/2}$. Thus, the probability becomes
\begin{equation}
P_{ide}^a=|T(p)T(q){\cal N}_a/{\cal N}_T^a|^2=\frac{1\pm |<\psi_T|\phi_T>|^2}{1\pm |<\psi|\phi>|^2}|T(p)T(q)|^2
\end{equation} 
and a similar expression for $P_{ide}^b$ with obvious modifications. For the mixture we have
\begin{equation}
P_{ide}^{mix}=\frac{1\pm |<\psi_T|\phi_T>|^2}{1\pm |<\psi|\phi>|^2}|aT(p)T(q)|^2+\frac{1\pm |<\varphi_T|\chi_T>|^2}{1\pm |<\varphi|\chi>|^2}|bT(\bar{p})T(\bar{q})|^2
\end{equation} 
Finally, following the same steps, the probability for a superposition of identical particles is
\begin{equation}
P_{ide}^{sup} = {\cal N}^2/{\cal N}_T^2
\end{equation}
with
\begin{eqnarray}
{\cal N}_T^{-2}=|aT_A(p)T_B(q)|^2({\cal N}_T^a)^{-2} +|bT_A(\bar{p})T_B(\bar{q})|^2({\cal N}_T^b)^{-2}+ 
\nonumber \\
4Re(a^*b T_A^*(p)T_B^*(q)T_A(\bar{p})T_B(\bar{q}) \times \nonumber \\
\{ <\psi_T|\varphi_T><\phi_T|\chi_T> \pm  <\psi_T|\chi_T><\phi_T|\varphi_T> \})
\end{eqnarray}
The probabilities depend on the coefficients of the superposition, the transmission coefficients and the overlapping between the one-particles states (both the initial and the transmitted ones). The coefficients of the superposition and the overlapping of the initial states are free parameters of the input states. In contrast, the transmission coefficients and the overlapping of the transmitted states must be evaluated . We do it in the next section.  

\section{The model.}
The transmission coefficient of a single mode or plane wave is well-known and can be found in many textbooks, for instance in \cite{gri}: 
\begin{equation}
T=\frac{2k_0k_1e^{-ik_0d}}{2k_0k_1 \cosh(k_1d)+i(k_1^2-k_0^2)\sinh(k_1d)}
\end{equation}
with $k_0 =(2mE/\hbar^2)^{1/2}$ and $k_1=(2m(V_0-E)/\hbar^2)^{1/2}$. The polar form of the coefficient is $T=|T|\exp(i\Omega _T)$, with
\begin{equation}
|T|^{-2}=1+\left(  \frac{k_1^2-k_0^2}{2k_1k_0} \sinh(k_1d) \right) ^2
\end{equation}
and
\begin{equation}
\Omega _T =arctg \left(  \frac{-(k_1^2 -k_0^2) \cos (k_0d) \sinh(k_1d)  - 2k_0k_1 \sin(k_0d) \cosh(k_1d)}{2k_0k_1 \cos(k_0d)\cosh(k_1d) -  (k_1^2 -k_0^2) \sin (k_0d)\sinh(k_1d) }    \right)
\end{equation}
As we have assumed multi-mode states sharply picked around the central value we can take, in a first approximation, the $T$ of the state as the transmission coefficient of the mode associated with the central value. A more exact evaluation of the transmission could be obtained via a multiple reflection expansion taking into account the multiple modes present in the wave packet.

In order to evaluate the overlap of the transmitted wave packets we recall that there is a time delay with respect to the incident ones, and that it is dependent on the energy . The explicit form of this delay is
\begin{equation}
\tau = \hbar \frac{d\Omega _T}{dE}
\end{equation} 
with the derivative evaluated at the central value of the packet \cite{gal}. This expression is usually denoted as the Wigner-Smith time \cite{wig,smi}.

The time spent by the pick of the free incident packet travelling a distance equal to the width of the barrier would be $\delta t_d = dm/p$. Because of the delay the transmission time through the barrier is $\delta t=\delta t_d +\tau$. We can introduce an effective momentum $p_e=md/\delta t$ that represents the momentum of a virtual free packet that would cross the barrier in the same time that the real particle.

The scalar product of the initial one-particle wave functions is $<\psi|\varphi>=\exp (-(p-\bar{p})^2/2P^2)$. As our model to pass the barrier is a virtual packet with the same form of the initial one but with an effective central momentum, the scalar product after the barrier will be 
\begin{equation}
<\psi_T|\varphi_T>=\exp (-(p_e-\bar{p}_e)^2/2P^2)
\end{equation}
and similar expressions for the other scalar products.  

\section{Results. }

We represent graphically the probabilities of double transmission in order to gain insight into the behaviour of the system. 
\begin{figure}[H]
\center
\includegraphics[width=8cm,height=7cm]{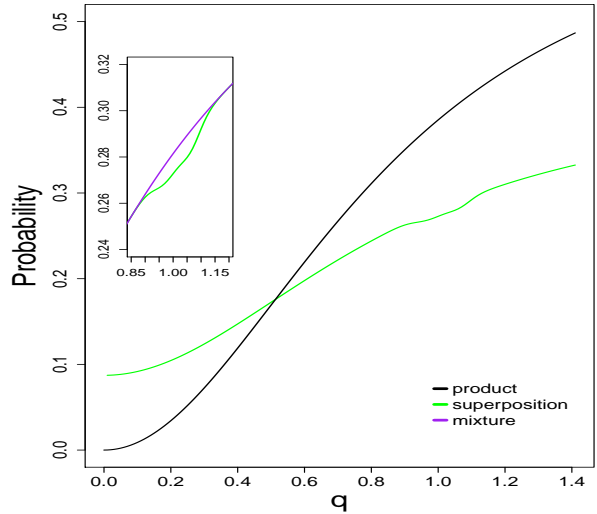}
\caption{Probability of double transmission versus the momentum $q$ in $m. u.$, for distinguishable particles. The inner figure contains a zoom of the comparison between the mixture and superposition curves, which only differ in the represented interval.}
\end{figure}
In the simulations we use an units system with the defining constants $m$, $\hbar$ and $V_0$, which are taken with numerical value one: $m=1$ (for the matter of simplicity, in the case of two distinguishable particles we take their mass equal) , $\hbar =1$ and $V_0=1$. This units system is a variant of the standard atomic one with the replacements of the defining length constant (the Bohr radius) by an energy one ($V_0$), and the electron mass by the mass of the particles present in the arrangement. Equivalently, we could use a system with a length constant, but in this case the Bohr radius should be replaced by the width of the barrier $d$, the natural length in the problem.  In our system the length unity, $l. u.$, and the momentum one, $m. u.$, are $l. u.=\hbar /(mV_0)^{1/2}$ and $m. u.=(2mV_0)^{1/2}$. In the representations we take   $p=0.95 m. u.$, $\bar{p}=1.05 m. u.$ and $\bar{q}=1 m. u.$ for the values of the central momenta,  $P=0.05 m. u.$ for the parameter of dispersion, and $d=0.7 l. u.$. On the other hand, $q$ is the independent variable of the problem, which as we are studying the regime $E<V_0$ runs in the interval $[ 0, \sqrt 2 )$. The rest of variables can be evaluated following similar steps. In particular, $\tau$ is a very lengthy expression that has not been explicitly included in the paper, but that can be deduced in a simple way via a standard derivative. 

First, we consider in Fig. 1 the case of distinguishable particles, where the effects of entanglement alone can be studied. We compare the probabilities for product, mixed and superposition states, that is, $P_{dis}^a$, $P_{dis}^{mix}$ and $P_{dis}^{sup}$. For the superposition coefficient we take the value $a=1/\sqrt 2$. We see that the curve of product states largely differs from those of superposition and mixture ones. For instance, for $q \approx 0$ the probability of double transmission for the superposition is not null, because the second term of the superposition contributes (${\bar p}$ and ${\bar q}$ are both different from zero). However, these differences are not related to physical processes. They only reflect the fact that we are comparing different states, a single multi-particle one with two-component multi-particle ones (the two components present in the superposition, either in a coherent or incoherent way). When inquiring about the physical processes present in the system, the proper comparison must be between the superposition and the mixture, containing both the two components of the full state. The comparison is presented in the inner box in the figure. We see in contrast that the differences are very small (for the rest of values not represented in the box both curves are almost equal). The difference between these two curves measures the influence of multi-particle coherence on the double transmission, which is small. As for pure states of distinguishable particles entanglement is the coherent superposition of multi-particle terms, we conclude that the entanglement influence on the transmission rates is small. From the physical point of view, the entanglement-based modifications can be understood as the quantum interference effects between the probability amplitudes associated with the two indistinguishable alternatives $|\psi>_A|\phi >_B \rightarrow |\psi_T>_A|\phi_T >_B$ and $|\varphi>_A|\chi >_B \rightarrow |\varphi_T>_A|\chi_T >_B$.      

Next, we study the effects of identity alone by comparing $P_{dis}^a$ and $P_{ide}^a$ in Fig. 2. The exchange effects only manifest when the overlap between the two particles is large, that is, for $q$ close to $p$. For the rest of values of $q$ the three curves are almost identical. The effects for bosons and fermions are clearly different, being more prominent in the second case. The Pauli principle is reflected in the figure as the interruption of the fermion curve for the point $q=p$, where the graphical program is unable to represent the indeterminacy $0/0$ present at that point in the probability distribution. The deviations with respect to the distinguishable case are due to the exchange effects, which, as it is well-known, can be interpreted as two-particle quantum interference effects between the indistinguishable alternatives $|\psi >_1|\phi >_2 \rightarrow |\psi_T >_1|\phi_T >_2$ and $|\phi>_1|\psi>_2 \rightarrow |\phi_T >_1|\psi_T >_2$. In this picture, entanglement and exchange effects have a common physical origin, quantum interference effects between indistinguishable alternatives.    
\begin{figure}[H]
\center
\includegraphics[width=8cm,height=7cm]{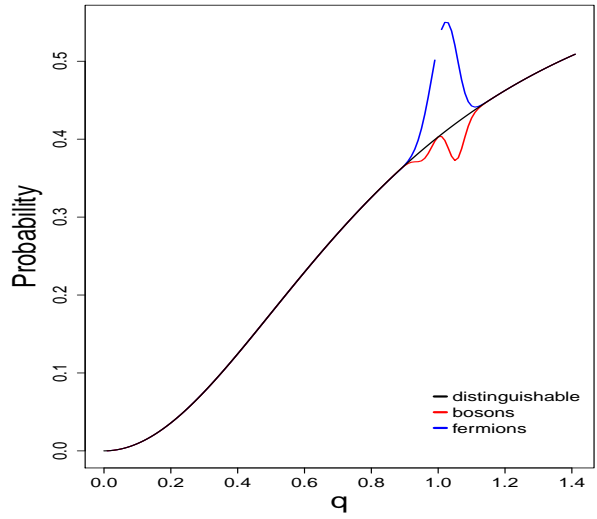}
\caption{As in Fig. 1, comparing distinguishable particles, bosons and fermions.}
\end{figure}  
\begin{figure}[H]
\center
\includegraphics[width=8cm,height=7cm]{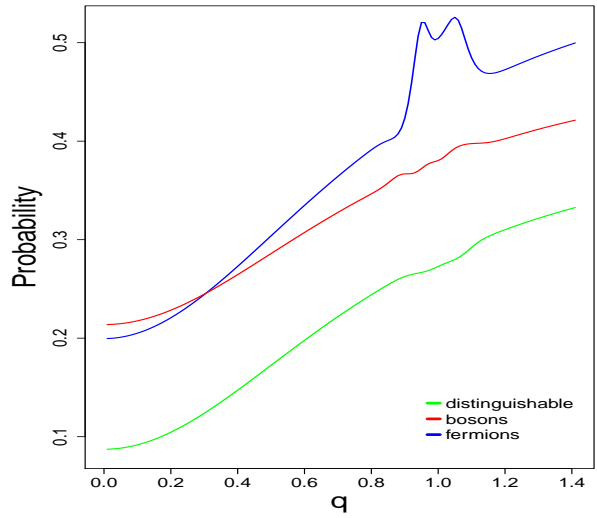}
\caption{As in Fig. 1, for the superposition of fermions, bosons and distinguishable particles. }
\end{figure}
Now, we move to the scenario where superposition and identity are present at once. We compare in Fig. 3 the double transmission for the superposition states of distinguishable and identical particles. The transmission rates are larger for identical particles, and for many intervals  of the variable momentum they even double the value of the distinguishable case. In the interval of large overlap between $p$ and $q$ the symmetrisation effects are more intense and we can observe sharper variations in the curves of identical particles. As the curve of distinguishable particles equals those of identical ones with the exchange interactions switched off, we conclude that the joint effects lead to rates much larger than those associated with superposition alone. The combined effects of superposition and symmetrisation can also be viewed as two-particle quantum interference effects. Now, we have a larger set of available alternatives for the double transmission, those associated with the terms $|\psi >_1|\phi >_2$, $|\phi>_1|\psi>_2$, $|\varphi >_1|\chi >_2$ and $|\chi>_1|\varphi>_2$ in the initial state. The strength and form of the interference effects are different from those found for other sets of alternatives. 
\begin{figure}[H]
\center
\includegraphics[width=8cm,height=7cm]{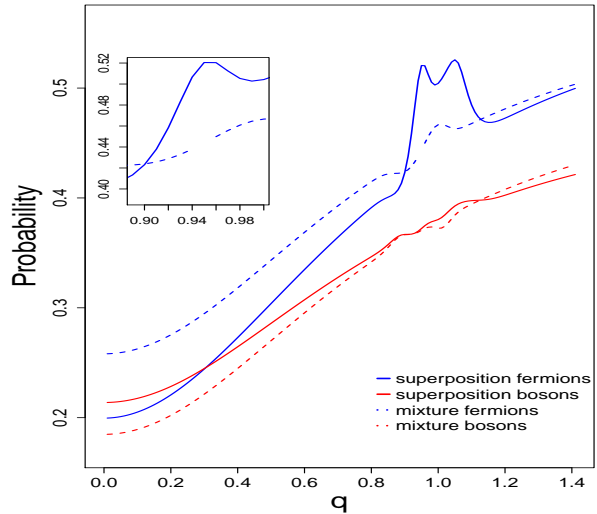}
\caption{As in Fig. 1 for $P_{ide}^{mix}$ and $P_{ide}^{sup}$ of bosons and fermions. The inner graphic is a zoom of the fermion case, showing a sharp image of the presence of an interruption of the curve in the case of the mixture, a feature that is absent in the superposition.}
\end{figure}
In Fig. 4 we compare $P_{ide}^{mix}$ and $P_{ide}^{sup}$ for both bosons and fermions in order to see that, as in the case of distinguishable particles, the differences between the mixture and the superposition are rather small (although not so small as in Fig. 1). Multi-particle coherence is not the dominant factor in the joint effects, being played that role by identity. The differences with a product state of identical particles (not represented here) are as in Fig. 1 very large, but as discussed above, these differences are not interesting when exploring the underlying physical mechanisms present in the problem. 

The most interesting aspect of Fig. 4 is the manifestation of fermion exclusion. The interruption associated with the exclusion is present for  $P_{ide}^{mix}$. The important point is that for the superposition the curve is continuous and smooth, reflecting the fact that the term that is excluded when is considered in an isolated way, can be present in a superposition without leading to any special effect. We shall consider this point in detail in the Discussion.

The above figures represent the behaviour of the system for the values detailed at the beginning of the section. Now, we discuss how the curves change for other choices of the parameters. When the coefficient of the superposition varies from $a=1/\sqrt 2$ to other values, we observe a progressive change of the curves towards the product state limits ($a=0$ and $a=1$), but without significant changes in the above results. For the distribution width $P$, we have that the differences between the mixture and  superposition curves (Figs. 1 and 4) and superposition ones (Fig. 3) become very small for values of $P$ close to zero (in this limit the scalar products present in the probabilities tend to zero), and increase when the parameter is larger than the value used here, although with similar patterns. Note that this parameter must be small because of the approximation of taking the transmission coefficient equal to that of the central value of the momentum  distribution, but not too small since then the overlaps, and consequently the exchange effects,  would be negligible. The only relevant modifications occur for low momenta values. When $p$, $\bar{p}$ or $\bar{q}$ are small the transmission rates largely decrease. Moreover, the separation between the distinguishable and identical curves of superposition (Fig. 3) are in general very small, just as between superposition and mixture ones (Figs. 1 and 4). The regime represented in the above four figures, that of large momentum values, is the most interesting because it provides the larger manifestations of the non-separability effects. In addition, the transmission rates increase and are easier to detect in potential tests of the problem.     

\section{Discussion.}

The graphics of the previous section illustrate the four main results derived in the paper. The first one is that in the regime of large incident momenta, the most interesting range of values in the problem because the non-separability effects increase with momentum, the combined effects lead to simultaneous transmission rates much larger than those associated with multi-particle superposition alone. The importance of multi-particle coherence in the problem can be analysed by comparing superposition states with mixtures of the same component states. The differences observed in that comparison are small, both for identical and distinguishable particles. The behaviour found here for tunnelling is by no means universal. In the case of two-particle two-slit interference of distinguishable systems \cite{yed}, the differences between the superposition and the mixed states are large, showing the importance of multi-particle coherence in that interferometric arrangement.    

The second result refers to the comparison of boson and fermion rates. For the interesting range of the momenta values, the differences between the boson and fermion cases can be as large as those between identical and distinguishable particles. The strength of the joint effects strongly depends on the bosonic or fermionic nature of the system.

The third result  reveals a new aspect of exclusion in entangled systems. In \cite{yap,yej} an example of excluded state not related to Pauli's principle was presented. Here, we have seen another manifestation of the relation between exclusion and multi-particle superposition: states that are forbidden by Pauli's principle can enter into a superposition state. This again highlights the fact that in the presence of entanglement only the two-fermion system can be in a definite pure state, whereas there are not individual pure states of the particles composing it. Then it is not possible to apply Pauli's principle to the component fermions, and there is not exclusion. In order to try an experimental demonstration of this property, it is clear that it is impossible to prepare the two terms and then to superpose them. Instead, we must prepare a two-fermion superposition of allowed states and then to change one of them. For instance, we can have a superposition of two two-electron states, one of them with the same spatial state for both electrons but with opposite orientations of the spin. Using electromagnetic pulses we can flip one of the spins (the intensity of the pulse must be low enough in order the probability of double flip to be negligible), evolving the above state to one with equal values of the spatial and spin variables.  

Finally, the fourth result relates to the understanding of the underlying physical mechanisms present in the problem. Non-separability processes are due to exchange and multi-particle superposition effects. We have noticed here that it is possible to have an unified view of these effects in terms of the quantum interference phenomenon. As it is well-known, both effects can be described as the interference between the indistinguishable alternatives available to the system. When both effects act at once the number of alternatives increases leading to different strengths and forms of the interference phenomena. Another important conceptual aspect of the problem is the dependence of the transmission rates on the overlap of the (initial and transmitted) one-particle states. This property is an immediate consequence of the dependence of the rates on the scalar products of these states. It is important to realize that there are two types of overlap. For identical particles we have overlap between the two particles, that is, an inter-particle phenomenon. This is a familiar property of exchange effects, which are only relevant for a large overlay between the two particles. On the other hand, the scalar products present in  the case of distinguishable particles represent the overlap between different states of the same particle, that is, they are related to the degree of non-orthogonality of these states. Non-orthogonality enters into the description of entanglement effects of distinguishable particles. In the case of identical particles in superposition, inter-particle ($<\psi|\phi>, <\varphi|\chi>,<\psi|\chi>$ and $<\phi|\varphi>$) and non-orthogonality ($<\psi|\varphi>$ and $<\phi|\chi>$) overlaps are simultaneously present.      

In summary, we have found that in the regime of large momenta there are large modifications of the simultaneous transmission rates due to the joint effects of entanglement and identity. The modifications depart from those associated with superposition and exchange effects alone. The comparison with other arrangements \cite{yap,yed} reveals that the combined effects of superposition and symmetrisation, and the influence of multi-particle coherence, do not show an universal behaviour, but they are dependent on the particular system analysed.

\end{document}